\newcommand{\algaas}{$\rm Al_{x}\-Ga_{1-x}\-As$} 
\newcommand{\gaasalgaas}{$\rm GaAs/Al_{x}\-Ga_{1-x}\-As$} 
 
\newcommand{\figcap}[1]{\caption{\protect \footnotesize #1}} 
\newcommand{\smalleps}[3]{\begin{figure}[#1]\centerline{ 
   \epsfxsize=6.9cm 
 \epsffile{#2}} 
   \figcap{#3} 
   \end{figure}} 
 
\documentstyle[twocolumn, epsf,emlines]{papers} 
 
\begin{document} 
 
\pagestyle{empty} 
\baselineskip 1em 
 
\begin{frontmatter} 
 
\title{ {\bf SHORT CIRCUIT CURRENT ENHANCEMENT IN \\ $\rm \bf
GaAs/Al_{x}Ga_{1-x}As$ MQW SOLAR CELLS} } 
 
\author[London]{James P. Connolly},
\author[London]{Keith W.J. Barnham}, 
\author[London]{Jenny Nelson},
\author[IRC]{Christine Roberts}, 
\author[Sheffield]{Malcolm Pate},
\author[Sheffield]{John S. Roberts}

\vspace{-0.1cm} 
\address[London]{Blackett Laboratory, Imperial College of Science, 
Technology and Medicine, London SW7 2BZ} 
\vspace{-0.16cm} 
\address[IRC]{IRC for Semiconductor Materials, Imperial College of Science, 
	 Technology and Medicine, London SW7 2BZ} 
\vspace{-0.1cm} 
\address[Sheffield]{ EPSRC III-V Facility, University of Sheffield, Sheffield S1 
3JD} 
 
\begin{abstract} 
 
\vspace{-3em} 
 \baselineskip=1.1em
ABSTRACT: The \gaasalgaas\ quantum well solar cell (QWSC) shows promise as a novel approach 
to higher efficiency solar cells but suffers from a poor short circuit current $J_{sc}$. We report on 
efforts to reduce this problem with the use of compositional grading and back surface mirroring. 
We present experimental quantum efficiency (QE) data on a range of compositionally graded 
QWSCs and devices in which the back surface of the cell is coated with a 
mirror, increasing the optical thickness of the quantum well layer in the long 
wavelength range. The experimental QE spectra are reproduced by a model which deals with 
arbitrary compositional profiles and optical cavities formed in the mirrored cells. The model is used 
to design an optimised QWSC, and projected $J_{sc}$ values given. Applications including II-VI and tandem solar cells are considered.
 
\vspace{0.2em}

\end{abstract} 
\end{frontmatter} 

\section{INTRODUCTION} 
 
\parindent 4ex 
 
The Quantum Well Solar Cell (QWSC) 
 was first proposed by Barnham and Duggan \cite{barnham90} to enhance efficiency by absorption of below bandgap radiation in the wells. This design 
shows an open circuit voltage ($V_{oc}$) greater than those expected of cells made from 
material with the quantum well effective bandgap \cite{barnham96}. The QWSC is therefore 
promising as an efficient solar cell design. 
 
The \gaasalgaas\  system \cite{green97} is an attractive material for the design of an 
QWSC since GaAs quantum wells in \algaas\ are unstrained, and  have been 
extensively studied for a wide range of optoelectronic devices. Previous work \cite{paxman93} has demonstrated a \gaasalgaas\ QWSC with an efficiency of 14.5\% AM1.5 and a  higher $V_{oc}$ than that of the highest efficiency GaAs device \cite{green97}. The overall efficiency remains 
low because of a low short circuit current density ($J_{sc}$). This is due to inefficient minority carrier transport in \algaas\, and the limit on the number of wells which can be incorporated. 
 
This work applies compositional grading techniques to address the transport issue, and light 
trapping to increase the quantum well absorptivity. 
 
\section{INCREASING  $\rm J_{sc}$} 
 
Figure \ref{QWSCFigure} shows a schematic of a graded, mirror backed QWSC. The back surface 
mirror causes light to perform several passes through the cell. This technique is well suited to the 
QWSC design since the wells convert light to current with nearly 100\% efficiency, but have a 
relatively low absorptivity.  
 
The compositionally graded p type emitter layer has a bandgap which increases towards the surface. 
Minority carriers generated in this region experience a pseudo-electric field due to the bandgap 
gradient and diffuse preferentially towards the junction. This increases minority carrier collection 
efficiency (\cite{konagai76} - \cite{hamaker85}).This technique also reduces the overall 
absorptivity of the emitter layer and  boosts the generation rate in the efficient intrinsic $i$ layer. 
 
The pseudo-electric field gradient increases $QE$ at short wavelengths, and the reduced emitter 
absorptivity increases $QE$ close to the bandgap $E_{g}$.
 
\begin{figure}[t] 
\begin{center} 
\unitlength 0.65mm
\linethickness{0.4pt}
\begin{picture}(106.10,92.12)
\put(60.58,56.00){\circle*{2.40}}
\put(44.45,27.06){\circle{3.40}}
\put(69.25,62.83){\circle*{2.40}}
\put(48.91,18.25){\circle{3.33}}
\put(60.41,21.92){\circle{3.33}}
\put(40.53,69.73){\line(0,-1){10.50}}
\put(40.53,59.23){\line(5,-1){7.67}}
\put(48.20,57.70){\line(0,1){9.53}}
\put(63.67,61.67){\line(4,-1){8.67}}
\put(56.33,64.17){\line(0,-1){10.50}}
\put(56.33,53.67){\line(5,-1){7.67}}
\put(64.00,52.13){\line(0,1){9.53}}
\put(72.33,59.42){\line(6,-1){5.67}}
\put(40.53,60.56){\line(1,0){7.67}}
\put(40.87,64.23){\line(1,0){7.67}}
\put(56.33,56.00){\line(1,0){7.67}}
\put(56.33,59.67){\line(1,0){7.67}}
\put(44.61,64.31){\circle*{2.40}}
\put(6.36,43.14){\makebox(0,0)[lc]{\small Incident light}}
\put(69.67,62.33){\vector(1,0){4.67}}
\bezier{36}(60.67,56.33)(63.33,58.67)(58.33,60.67)
\bezier{28}(58.33,60.67)(57.00,63.00)(61.00,62.67)
\bezier{32}(61.00,62.67)(66.00,61.33)(68.67,62.33)
\put(63.67,16.67){\line(4,-1){8.67}}
\put(72.36,14.52){\line(6,-1){5.67}}
\put(63.67,16.67){\line(0,1){6.67}}
\put(63.33,23.67){\line(-3,1){6.67}}
\put(56.67,19.00){\line(0,1){6.67}}
\put(47.87,22.23){\line(0,1){6.67}}
\put(47.53,29.23){\line(-3,1){6.67}}
\put(40.87,24.56){\line(0,1){6.67}}
\put(40.87,26.90){\line(1,0){7.00}}
\put(56.67,22.00){\line(1,0){7.00}}
\bezier{40}(60.67,20.00)(61.67,14.67)(57.67,17.33)
\bezier{16}(57.67,17.33)(54.33,17.67)(54.00,16.67)
\bezier{28}(54.00,16.67)(53.67,14.33)(50.67,17.33)
\put(47.33,18.33){\vector(-1,0){3.67}}
\put(59.28,3.00){\makebox(0,0)[cc]{\small i}}
\put(88.33,3.00){\makebox(0,0)[cc]{\small n}}
\put(85.33,3.00){\vector(-1,0){5.33}}
\multiput(40.55,69.62)(-0.31,0.12){17}{\line(-1,0){0.31}}
\put(35.33,71.60){\line(0,1){0.00}}
\multiput(40.83,24.45)(-0.31,0.12){17}{\line(-1,0){0.31}}
\put(35.61,26.43){\line(0,1){0.00}}
\multiput(35.61,26.43)(-0.83,0.11){9}{\line(-1,0){0.83}}
\put(78.00,58.48){\line(1,0){12.66}}
\multiput(48.14,67.24)(0.27,-0.11){8}{\line(1,0){0.27}}
\multiput(51.40,66.04)(0.25,-0.11){5}{\line(1,0){0.25}}
\multiput(53.47,65.22)(0.26,-0.11){4}{\line(1,0){0.26}}
\multiput(56.35,64.12)(-0.25,0.10){4}{\line(-1,0){0.25}}
\multiput(47.91,22.18)(0.27,-0.11){8}{\line(1,0){0.27}}
\multiput(51.17,20.98)(0.25,-0.11){5}{\line(1,0){0.25}}
\multiput(53.24,20.16)(0.26,-0.11){4}{\line(1,0){0.26}}
\multiput(56.12,19.05)(-0.25,0.10){4}{\line(-1,0){0.25}}
\multiput(35.35,71.63)(-0.39,0.12){45}{\line(-1,0){0.39}}
\put(96.33,69.67){\makebox(0,0)[cb]{\tiny Mirror}}
\put(77.37,3.06){\vector(1,0){0.2}}
\put(63.40,3.06){\line(1,0){13.97}}
\put(32.40,3.27){\vector(-1,0){0.2}}
\put(54.67,3.27){\line(-1,0){22.27}}
\put(29.67,3.33){\vector(1,0){0.2}}
\put(24.67,3.33){\line(1,0){5.00}}
\put(4.28,89.47){\circle*{2.40}}
\put(22.28,78.31){\circle*{2.40}}
\put(35.67,73.67){\vector(3,-1){0.2}}
\multiput(23.00,77.67)(0.37,-0.12){34}{\line(1,0){0.37}}
\put(4.25,24.58){\circle{3.33}}
\bezier{20}(4.39,89.80)(3.72,87.47)(6.05,86.14)
\bezier{16}(6.05,86.14)(7.72,85.47)(8.05,83.14)
\put(4.34,83.00){\vector(0,1){0.2}}
\put(4.34,26.33){\line(0,1){56.67}}
\put(8.70,80.70){\vector(1,-2){0.2}}
\bezier{16}(7.95,83.16)(7.20,81.45)(8.70,80.70)
\bezier{96}(7.30,49.72)(15.89,58.31)(24.48,50.37)
\bezier{76}(24.48,50.37)(29.64,45.43)(41.02,49.94)
\put(41.28,49.97){\vector(4,1){0.2}}
\put(40.90,49.88){\line(1,0){0.38}}
\bezier{88}(88.01,39.28)(79.21,35.84)(66.75,39.49)
\bezier{76}(66.75,39.49)(60.52,41.64)(48.28,39.49)
\put(47.55,39.50){\vector(-1,0){0.2}}
\put(48.28,39.50){\line(-1,0){0.73}}
\put(87.33,31.00){\makebox(0,0)[rc]{\tiny reflected}}
\put(8.34,87.66){\makebox(0,0)[lc]{\tiny thermalisation}}
\put(25.67,79.33){\makebox(0,0)[lc]{\tiny drift/diffusion}}
\put(2.66,3.33){\makebox(0,0)[lc]{\small graded p}}
\multiput(91.08,9.15)(0.41,0.12){26}{\line(1,0){0.41}}
\multiput(91.08,11.15)(0.41,0.12){26}{\line(1,0){0.41}}
\multiput(91.08,13.15)(0.43,0.12){25}{\line(1,0){0.43}}
\multiput(91.08,15.15)(0.43,0.12){25}{\line(1,0){0.43}}
\multiput(91.08,17.15)(0.41,0.12){26}{\line(1,0){0.41}}
\multiput(91.08,19.15)(0.41,0.12){26}{\line(1,0){0.41}}
\multiput(91.08,21.15)(0.41,0.12){26}{\line(1,0){0.41}}
\multiput(91.08,23.15)(0.41,0.12){26}{\line(1,0){0.41}}
\multiput(91.08,25.15)(0.41,0.12){26}{\line(1,0){0.41}}
\multiput(91.08,27.15)(0.41,0.12){26}{\line(1,0){0.41}}
\multiput(91.08,29.15)(0.43,0.12){25}{\line(1,0){0.43}}
\multiput(91.08,31.15)(0.43,0.12){25}{\line(1,0){0.43}}
\multiput(91.08,33.15)(0.41,0.12){26}{\line(1,0){0.41}}
\multiput(91.08,35.15)(0.41,0.12){26}{\line(1,0){0.41}}
\multiput(91.08,37.15)(0.41,0.12){26}{\line(1,0){0.41}}
\multiput(91.08,39.15)(0.41,0.12){26}{\line(1,0){0.41}}
\multiput(91.08,41.15)(0.41,0.12){26}{\line(1,0){0.41}}
\multiput(91.08,43.15)(0.41,0.12){26}{\line(1,0){0.41}}
\multiput(91.08,45.15)(0.41,0.12){26}{\line(1,0){0.41}}
\multiput(91.08,47.15)(0.41,0.12){26}{\line(1,0){0.41}}
\multiput(91.08,49.15)(0.41,0.12){26}{\line(1,0){0.41}}
\multiput(91.08,51.15)(0.41,0.12){26}{\line(1,0){0.41}}
\multiput(91.08,53.15)(0.41,0.12){26}{\line(1,0){0.41}}
\multiput(91.08,55.15)(0.41,0.12){26}{\line(1,0){0.41}}
\multiput(91.08,57.15)(0.41,0.12){26}{\line(1,0){0.41}}
\multiput(91.08,59.15)(0.41,0.12){26}{\line(1,0){0.41}}
\put(90.89,9.07){\framebox(10.92,53.19)[cc]{}}
\put(28.24,27.43){\line(-1,0){27.00}}
\multiput(17.93,76.87)(-0.62,0.12){26}{\line(-1,0){0.62}}
\put(78.05,13.55){\line(1,0){12.70}}
\put(0.00,0.00){\framebox(106.10,92.12)[cc]{}}
\end{picture}
\figcap{Schematic of a graded and mirror backed showing generation, 
thermalisation and drift in the graded p layer\label{QWSCFigure}} 
\end{center} 
\end{figure} 
 
\section{LIGHT INTENSITY IN A QWSC\label{lightsection}} 
 
The light intensity in a QWSC with a back mirror (figure \ref{QWSCFigure}) is determined by 
calculating the electric field strength in the solar cell as a function of position and wavelength, for 
position dependent optical parameters \cite{connolly97}. The parameters used are defined in Table \ref{table1}. The total normalised light amplitude $\cal E$ inside the cell is found by summing 
successive reflections from front and back surfaces to infinity, and can be expressed as: 
\begin{equation} 
\label{sumlimit} 
{\cal E}         =          \left(\frac{ 
				e^{-\int_{0}^{x}kdx'} 
			+\frac{r_{b}e^{-2\int_{0}^{d}kdx'}} {e^{-
\int_{0}^{x}kdx'}} 
				    } 
				    { 
1-r_{f}r_{b}e^{-2\int_{0}^{d}kdx'} 
				     }\right) 
\end{equation} 
The light intensity in the solar cell is then given by 
\begin{equation} 
\label{amplitudetointensity} 
{\cal I}(x,\lambda)={\cal I}_{i}(1-R)\left({\cal E E}^{*}\right) 
\end{equation} 
and hence the generation rate at any position is 
\begin{equation} 
\label{generationrate} 
G(x,\lambda)= \alpha(x,\lambda){\cal I}(x,\lambda) 
\end{equation} 
 
\begin{table}[t] 
\begin{center} 
\small 
\begin{tabular}{| l | c |} \hline 
{\bf Parameter}      &         {\bf Meaning}                                                            \\ 
\hline
$\lambda$      &        {\tiny wavelength  }                                                            \\ 
\hline  
$\alpha$         &        {\tiny absorption  }                                                            \\  
		       &        {\tiny coefficient \cite{aspnes89},\cite{paxman93}  }      \\ 
\hline  
$n$                 &        {\tiny refractive index \cite{aspnes89}  }                         \\ 
\hline 
$k$                 &       {\tiny complex wavevector  }                                              \\ 
		       &      {\tiny $=\frac{\alpha}{2} - \frac{2\pi i n}{\lambda}$   }   \\ 
\hline 
$R$          &        {\tiny measured reflectivity }                                                                    \\  
\hline 
$r_{f}$          &       {\tiny front surface complex      }                                      \\  
		       &       {\tiny amplitude reflectivity     }                                        \\ 
\hline 
$r_{b}$          &       {\tiny back surface complex  }                                         \\  
			&      {\tiny amplitude reflectivity }                                            \\ 
\hline 
\end{tabular} 
\figcap{Parameters needed by the light intensity calculation \label{table1}}  
\end{center} 
\end{table} 
 
\section{PHOTOCURRENT AND QE MODEL\label{model}} 
 
The photocurrent in the charge neutral sections of the emitter and base layers is calculated by 
solving minority carrier current and continuity equations with position 
dependent parameters \cite{connolly97}. The equation takes the following form 
\begin{equation} 
\label{differentialeq} 
		 \frac{d^{2}n}{dx^{2}} 
		 +b(x)\frac{dn}{dx} 
		 +c(x) n+g=0  
\end{equation} 
where $n$ is the minority carrier concentration and the parameters are defined in Table \ref{table2}. This equation is solved numerically using standard second order numerical methods. The boundary conditions are $n=0$ at the depletion edge in the low 
injection depletion approximation, and a surface recombination current given by the surface 
recombination velocity \cite{hovel75}. The photocurrent from the $p$ and $n$ layers is then 
 given by the gradient of the minority carrier concentration at the $p-i$ and $i-n$ interfaces respectively. In the depletion approximation, all carriers generated in the $i$ layer are collected. The current from the $i$  layer is then simply the integral of the generation rate across it.

The photocurrent density $J_{ph}$ is given by the sum of $p$,$i$ and $n$ layer photocurrent 
contributions. The $QE$ of the cell is then
\begin{equation} 
\label{QEdef} 
QE(\lambda)=\frac{j_{ph}(\lambda)}{q{\cal I}_{i}(\lambda)} 
\end{equation} 
 
\begin{table}[t] 
\begin{center} 
\small 
\begin{tabular}{| l | c |} \hline 
{\bf Parameter }     &      {\bf Meaning }
\\ \hline
$X$      &        {\tiny aluminium mole fraction}                                                          
\\ \hline  
$D$      &        {\tiny diffusivity \cite{hamaker85}}                                                     
\\ \hline  
$L$     & {\tiny diffusion length 
\cite{hamaker85},\cite{ahrenkiel91},\cite{ahrenkiel92}}  \\  \hline 
$E$      &        {\tiny effective electric field 
\cite{konagai76},\cite{hutchby76},\cite{sassi83}}\\ 
	     &        {\tiny $= \nabla_{\!x}E_{g}$ }                                                               
\\ \hline 
$S$      &        {\tiny surface recombination velocity \cite{timmons92}}                      
\\ \hline 
${\cal E}_{f}$   &  {\tiny normalised effective field $=qE/(k_{B}T)$ }
\\ \hline 
${\cal D}_{X}$   &  {\tiny normalised diffusivity field  
$=\nabla_{\!X}D/D$ } \\ \hline  
{\normalsize $b$ }   &    {\tiny field-like parameter  =  ${\cal E}_{f}+{\cal 
D}_{X}$ }\\ \hline 
{\normalsize $c$  }   &    {\tiny diffusion length-like parameter }\\ 
		  &   {\tiny $={\cal E}_{f}{\cal D}_{X} + \nabla_{\!x}{\cal E}_{f}-
L^{-2}$ }\\ \hline 
{\normalsize $g$}           &   {\tiny normalised generation rate } \\  
		  &   {\tiny $=G/D$  }\\ \hline  
\end{tabular} 
\figcap{Definitions of model parameters  with references \label{table2}}  
\end{center} 
\end{table} 
 
\section{EXPERIMENTAL RESULTS} 
 
\subsection{Samples} 
Sample were grown by MBE at Imperial College and by MOVPE at the 
University of Sheffield III-V Facility.  30 and 50 well QWSCs with and without compositional grades were grown. Control samples are identical in structure to the QWSCs except that the well material is replaced with \algaas\ with the same composition as the barrier. The wafers were processed into test devices at Sheffield. These are circular 1mm diameter mesa structures with a 0.545mm diameter optical window. 
 
Samples with back surface mirrors were processed by etching the GaAs substrate down to the back 
of the QWSC structure under the optical window. The back surface of the n region acts as an etch 
stop. The resulting pit is then coated with a metallic mirror. 
 
\subsection{Modelling} 
The model is used to reproduce experimental data. The model parameters listed in Table 
\ref{table2} are determined by growth. The QE of samples without compositional grades is 
sensitive to the diffusion length $L$ at wavelengths near the emitter band edge, and to $\cal S$, the 
ratio of surface recombination velocity and diffusivity, at short wavelengths. Therefore, we use $L$ 
and $\cal S$ as independent fitting parameters.

Modelling shows that graded samples are relatively insensitive to diffusion length. They are highly 
sensitive, however, to the parameter ${\cal D}_{X}$ defined in table \ref{table2}, which reflects how rapidly the diffusivity $D$ varies with Al fraction. In the absence of direct measurements of $D$ as a function of composition, we assume an exponential dependence \cite{connolly97}of $D$ on $X$, as recommended by Hamaker \cite{hamaker85}, which reduces ${\cal D}_{X}$ to a constant. The values needed to reproduce the data are similar or better  than those reported by Hamaker.

Comparison with published values gives us an indication of the efficiency of minority carrier 
transport in our material. We find that our values are lower than published numbers at aluminium 
fractions above about 0.3, indicating worse minority carrier transport. 
 
\smalleps{htbp}{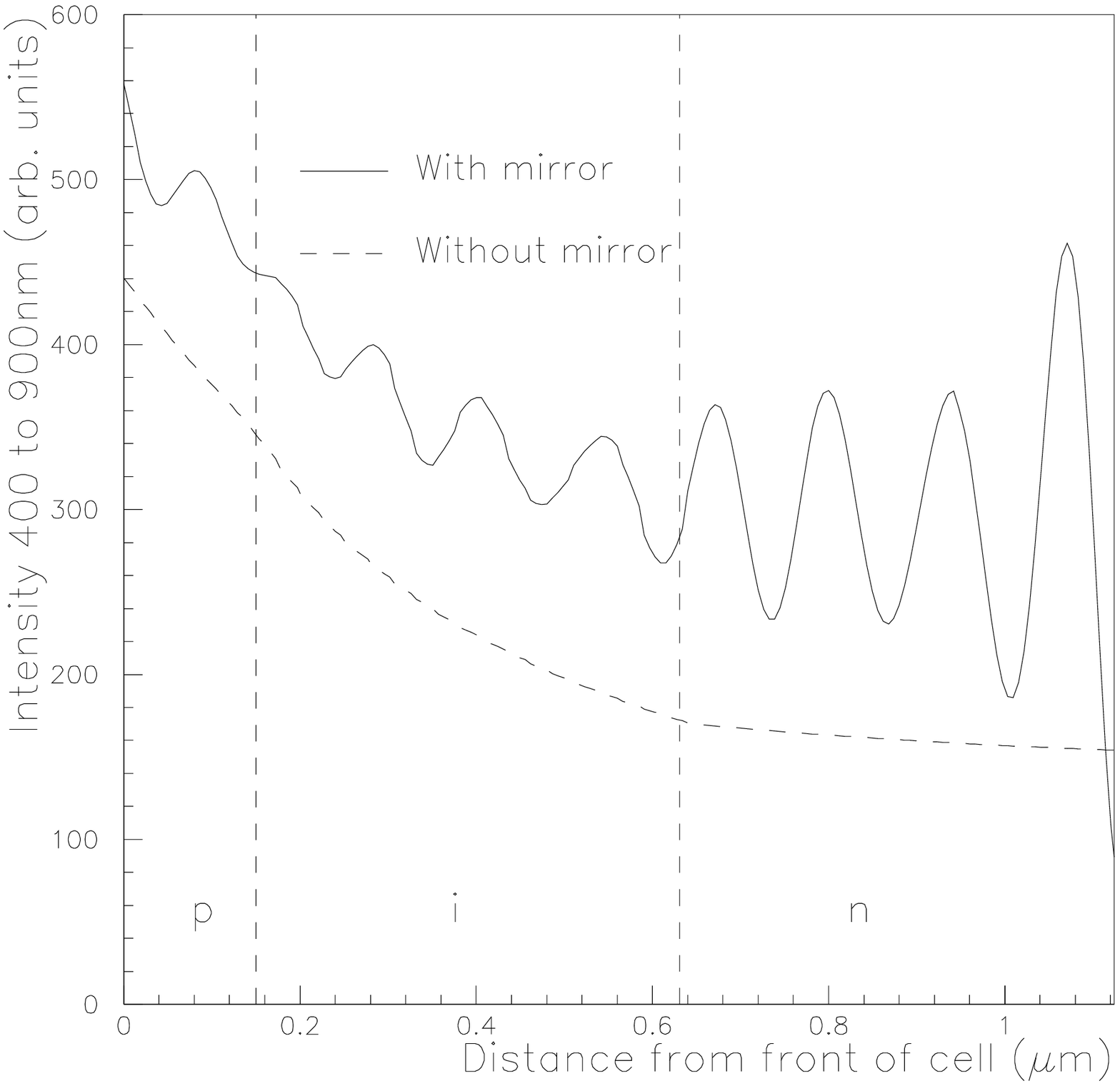} 
{Total light intensity 400-900nm in a QWSC with and without a mirror \label{lightfig}} 
 
\smalleps{htbp}{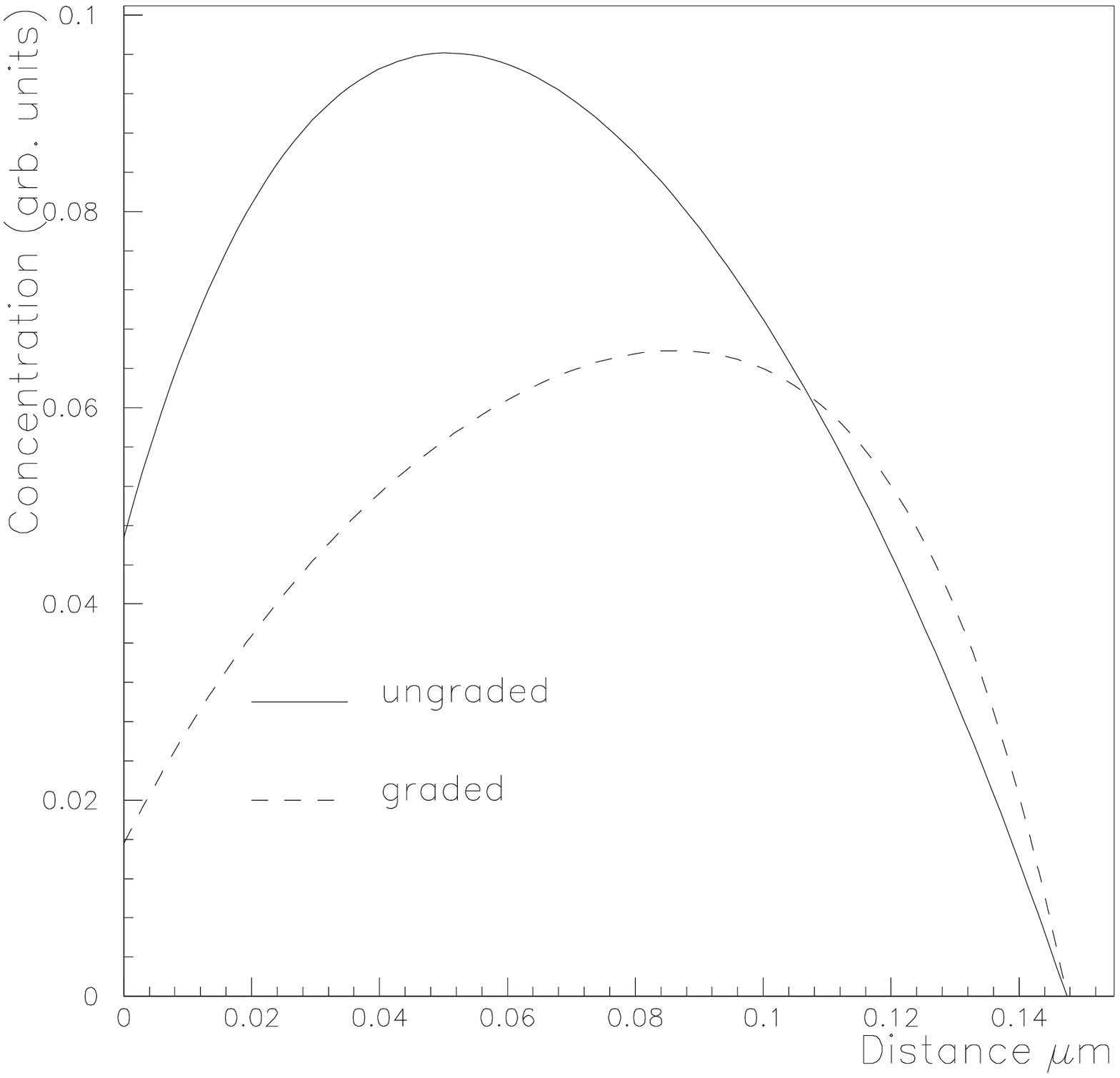} 
{Minority carrier concentration in a p type emitter with and without a compositional grade \label{carriersfig}} 
 
\subsection{Results}
Figure \ref{lightfig} shows the panchromatic increase in light levels in a QWSC structure with and without a mirror. Fabry-Perot oscillations are visible in the mirrored case. Figure \ref{carriersfig} shows minority carrier profiles in a 0.3 Al fraction p type emitter and in an identical emitter with an Al fraction graded from 0.4 to 0.3. We note that the gradient of the minority carrier concentration is decreased at the surface and increased at the p - i interface. The former implies a reduced surface recombination, and the latter an enhanced $J_{sc}$.

Figure \ref{mirrorfig} shows the experimental QE of the an ungraded 30 well QWSC with and 
without a back surface mirror. Also shown is the model for this cell, showing good agreement. The 
experimental short circuit current enhancement under an AM1.5 spectrum is 27\% in this case. 
 
Figure \ref{4034fig} shows experiment and modelling for a 30 well QWSC with a base aluminium 
fraction of  0.30 graded to 0.67  over 0.15 $\mu$m, also showing a good fit. The $J_{sc}$ enhancement due to the grade is 24\%. 
 
This gives an combined overall $J_{sc}$ increase of over 50\% extra $J_{sc}$ for a 30 well QWSC with both design improvements. The percentage enhancement due to the mirror decreases as more wells are introduced and the mirror becomes redundant. Similarly, the enhancement due to the grade decreases as the material quality is improved, and the grade also becomes unnecessary.

\smalleps{htbp}{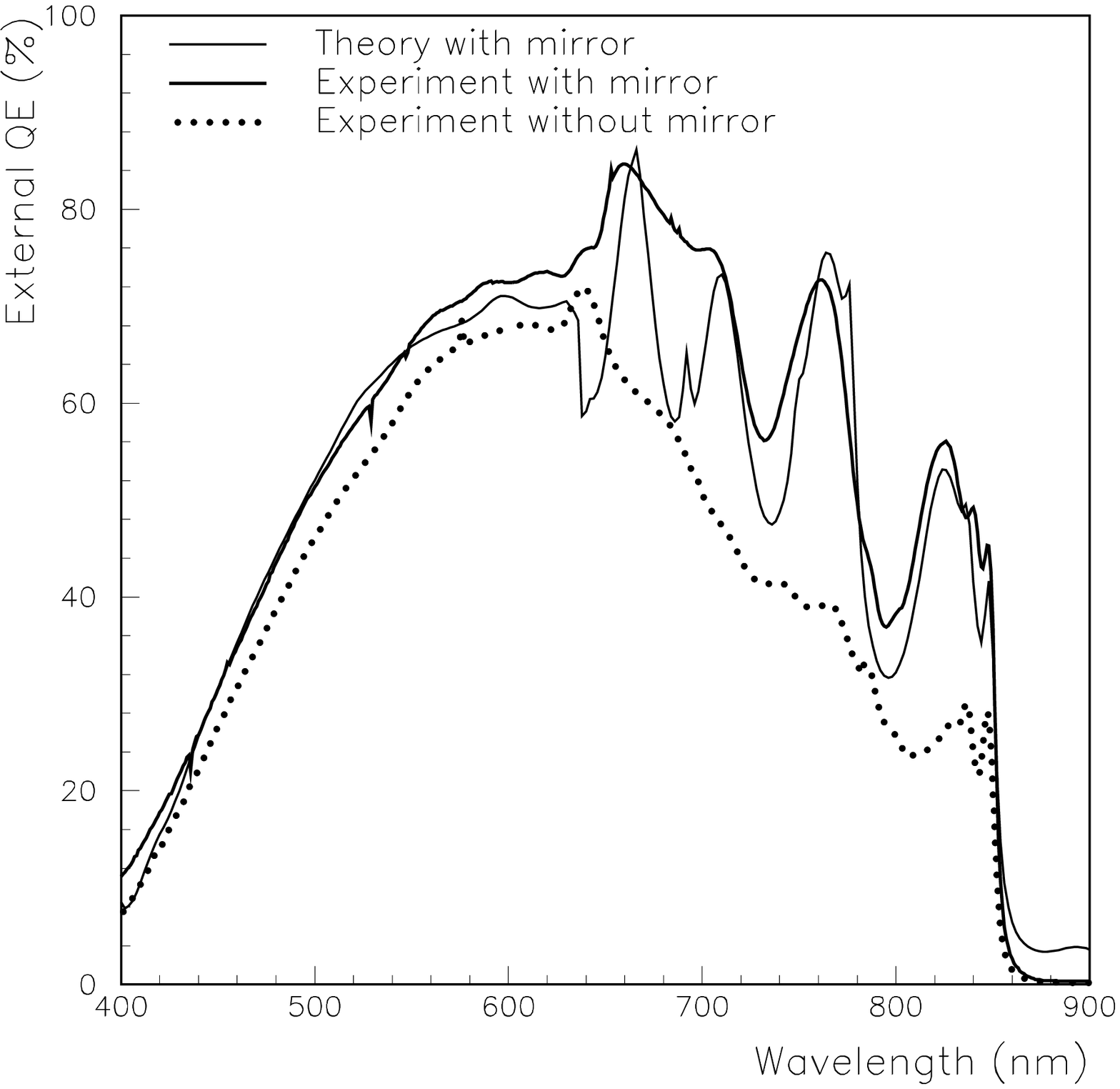} 
{QE of a 30 well QWSC with and without mirror, together with theory for the mirrored case 
\label{mirrorfig}} 
 
\smalleps{htbp}{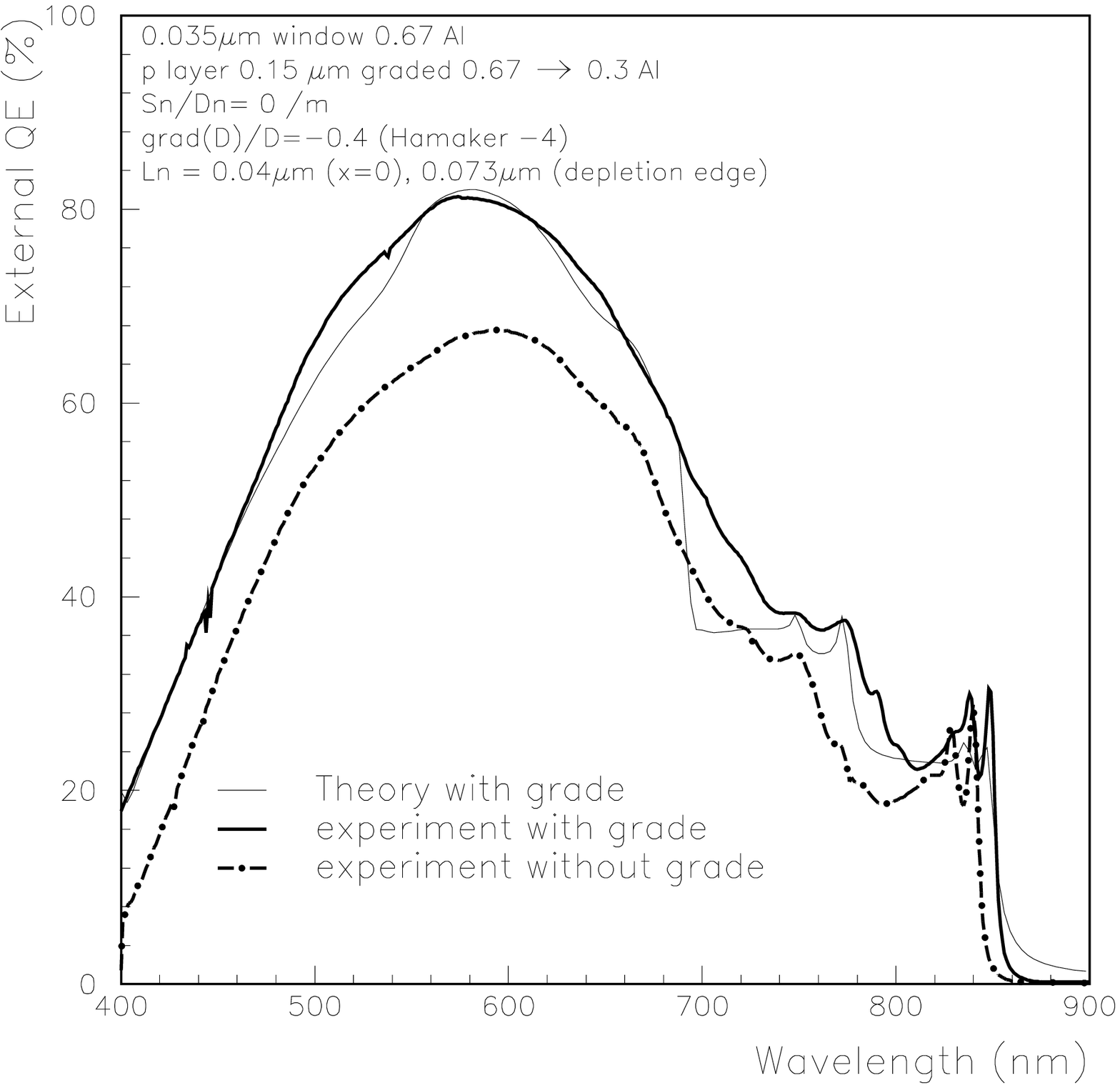} 
{Experimental data and theoretical fit for a 30 well QWSC with a base aluminium fraction of 0.30 
graded to 0.67  over 0.15 $\mu$m \label{4034fig}} 
 
\section{Optimised QWSCs} 
 
\subsection{Present Material} 
Having separately established the viability of both techniques for enhancing the $J_{sc}$, we use 
the model to predict an optimised design. The $p$ layer is thinned to $0.1\mu m$ in order to reduce losses in the $p$ layer without compromising the series resistance characteristics. The base 
aluminium fraction is set at 0.30, for which an optimum of 0.40 at the top of the grade is found. 
Transport parameters are consistent with those observed in present material. 
 
Experience shows that 50 wells can safely be incorporated into the $i$ region without 
compromising the voltage performance of the QWSC. Modelling of this cell is shown in figure 
\ref{optfig}, and produces a short circuit current increase of 39\% compared to a suitable 50 well QWSC control with a standard 0.15$\mu$m p layer, no grade, and no mirror. The efficiency prediction for  the optimised structure is between 19\% and 20\%, using the dark current of a 50 well QWSC.
 
\smalleps{htbp}{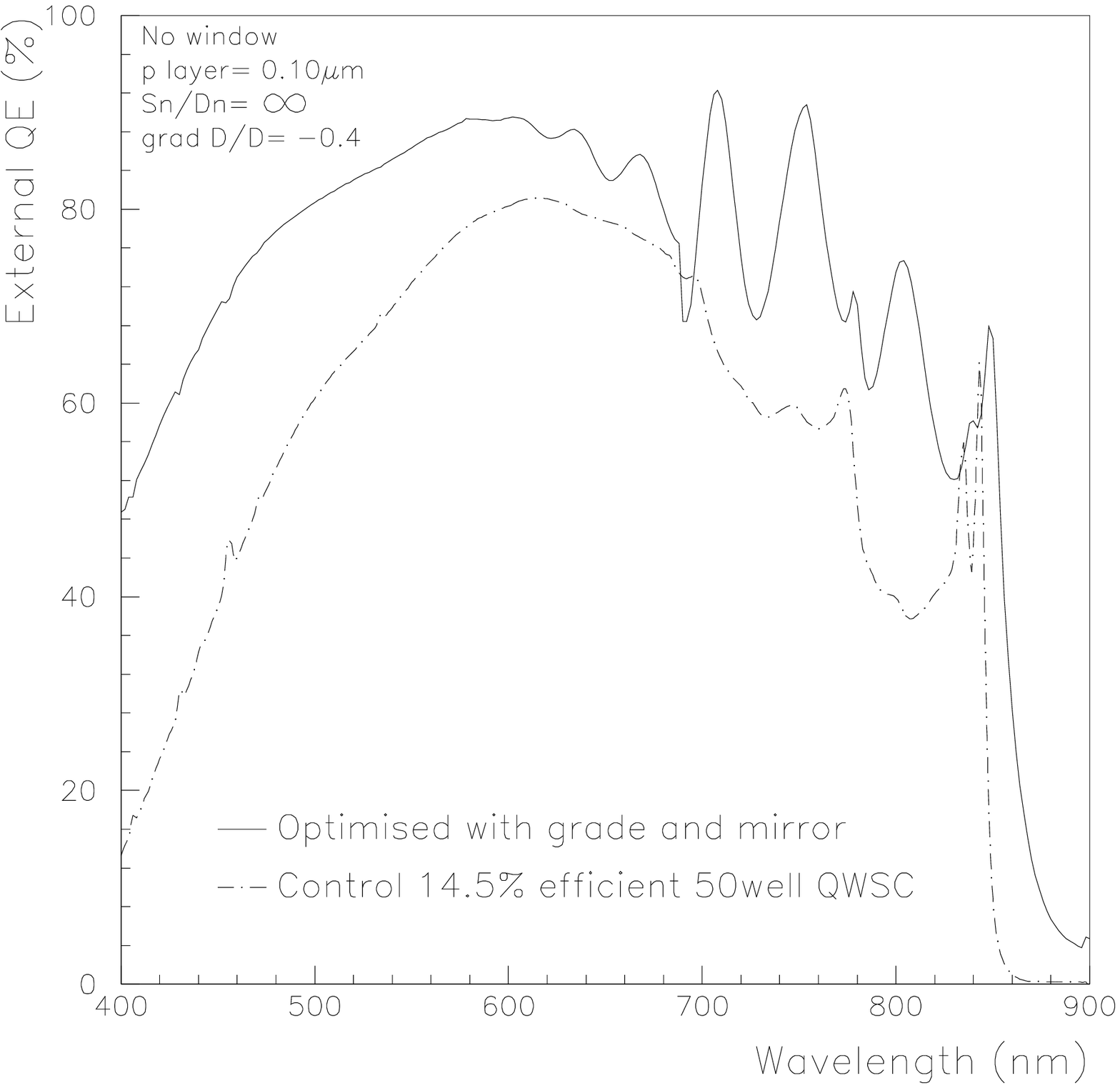} 
{Comparison of experimental QE for 14.5\% AM1.5 efficient QWSC and theoretical projection for the optimised device with grade over 0.01 $\mu m$ and back surface mirror \label{optfig}} 
 
\subsection{Ideal Material} 
Further efficiency enhancements are possible in material with improved minority transport 
parameters. Ludowise \cite{ludowise84} reports very high diffusion lengths in a series of solar cell 
samples grown by MOVPE. These values were derived from internal $QE$ spectra, using a method 
similar to the one used by us. 
 
An optimisation based on the Ludowise diffusion length of $L_{n}\sim 1.4\mu m$ at Al fraction 
$0.30$ yields a value of  $J_{sc}=25.6 \rm mA cm^{-2}$. This corresponds to an efficiency 
between $21\%$ and $24\%$ for low and high values of $F\!F$ and $V_{oc}$ in AM1.5. These 
results apply to a 50 well sample with a back surface mirror, a $0.02\mu m$ window of 
composition $0.9$ Al, no compositional grade and  an optimum p layer thickness of $0.15 \mu m$. 
 
\section{Further Applications} 
Similar techniques may be applied to other materials which suffer from minority carrier transport 
problems such as {$\rm Cd\-Hg\-Te$} or {$\rm CuIn\-Ga\-Se$}. 

The short circuit current of a QWSC may be easily tailored by changing the number and/or width of 
the quantum wells, without affecting the p and n layers. This means that the design of the 
conventional part of the cell need not be compromised by current matching requirements. We 
therefore propose another application which is the tandem QWSC. Ideally, this would use two QWSCs, but here we present a preliminary example in \gaasalgaas.

The QE of a structure optimised for AM1.5 is shown in figure \ref{tandemfig}. The top cell is based on previous high Al fraction samples. It has a 0.075$\mu m$ p type emitter graded from 0.67 to 0.48 Al and an 0.8 $\mu m$ $i$ layer. The bottom cell is a standard high efficiency GaAs design. The QE and $J_{sc}$ values compare favourably with the tandem world record holder, a cell reported by Takamoto \cite{takamoto96}. Furthermore, we expect higher voltages \cite{barnham96}.Excluding shading, we find that  41 wells of width 31$\AA$ are sufficient to match currents at 13.6$mA\- cm^{-2}$. Under AM0, currents are matched simply by reducing the number of wells to 33 for a $J_{sc}$ of 16.0$mA\- cm^{-2}$.

\smalleps{htbp}{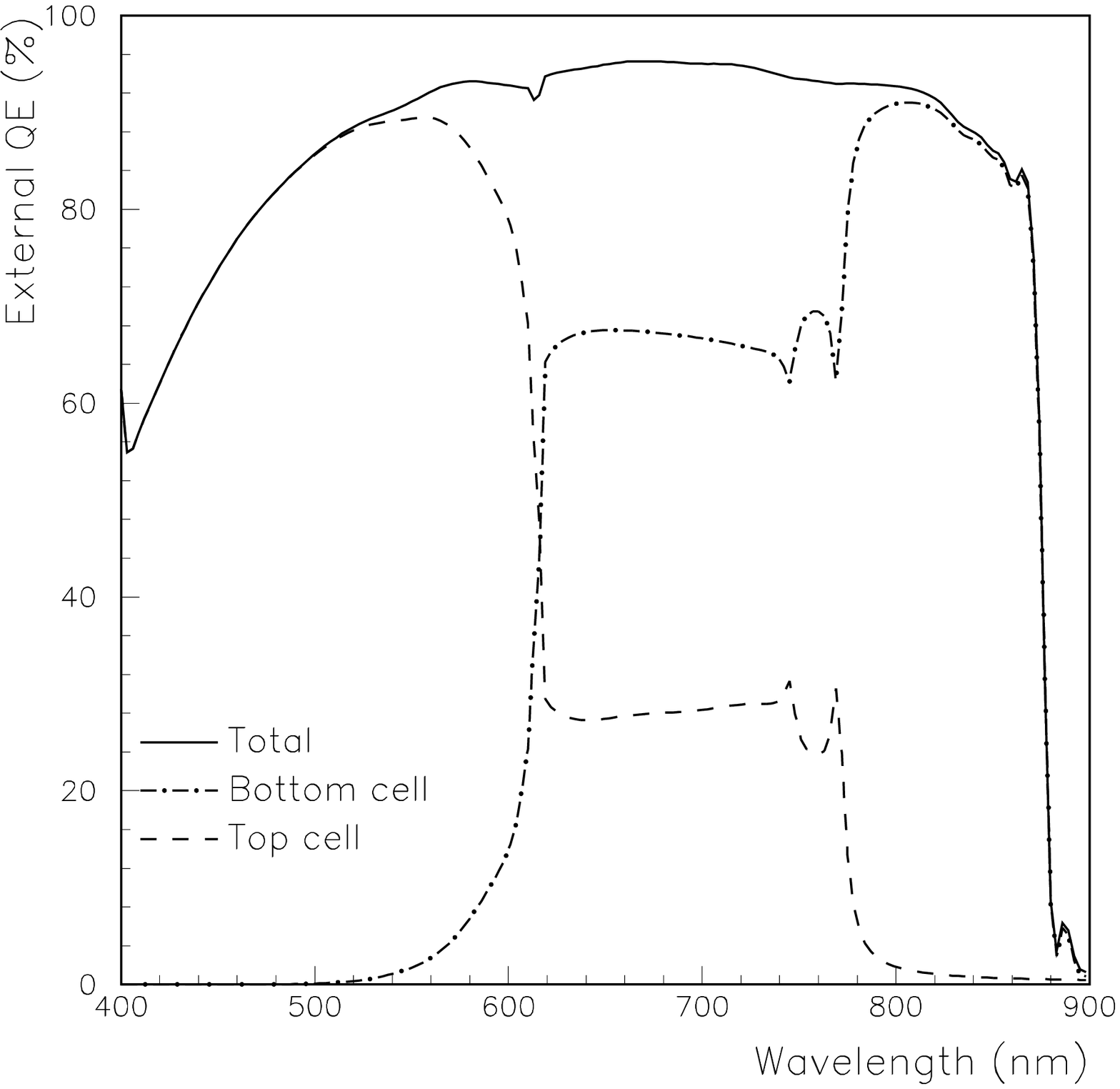} 
{Illustrative example of a QWSC tandem \label{tandemfig}} 
 
\section{Conclusion} 
Both compositional grading and back surface mirror techniques can significantly increase the 
$J_{sc}$ of an \algaas\ QWSC, and is particularly suited to poor material. In general, $J_{sc}$ optima are found at smaller grades as the emitter efficiency improves. Theory shows that mirrors help QWSCs made from efficient material, but that grades are unnecessary in this case. We predict efficiencies of up to 20\% with existing material, or 24\% with optimum material.
 
We conclude that the compositional grading technique is of interest for QWSCs, and solar cells 
in general which are made from material with low minority carrier transport efficiency. They may furthermore be attractive for tandem applications both because of potentially higher voltages, and because of the simplified current matching properties.

\ack
We are grateful to the Greenpeace Trust, who made this work possible, and to the EPSRC.

\end{document}